\documentclass[12pt]{article}

\usepackage{amsmath,amssymb}
\usepackage{type1cm}

\numberwithin{equation}{section}

\begin{document}

\begin{titlepage}
\renewcommand{\thefootnote}{\fnsymbol{footnote}}
\vspace*{-5mm}
\hfill
\vbox{
    \halign{#\hfil         \cr
           CERN-PH-TH-2009-183 \cr
           OU-HET 642 \cr
           RIKEN-TH-169 \cr
           } 
      }  
\vspace*{5mm}
\begin{center}
{\Large {\bf 
Three-flavor quark mass dependence of \\ baryon spectra in holographic QCD
} }

\vspace*{15mm}
{\sc Koji Hashimoto}$^{a}$\footnote{e-mail: {\tt koji@riken.jp}},
{\sc Norihiro Iizuka}$^{b}$\footnote{e-mail: {\tt norihiro.iizuka@cern.ch}},
{\sc Takaaki Ishii}$^{a,c}$\footnote{e-mail: {\tt ishiitk@riken.jp}},
{\sc Daisuke Kadoh}$^{a}$\footnote{e-mail: {\tt kadoh@riken.jp}}

\vspace*{7mm} 
{\it {$^{a}$ Theoretical Physics Laboratory, Nishina Center, RIKEN, 
Saitama 351-0198, Japan
}}\\ 

\vspace*{5mm}
{\it {$^{b}$ 
Theory Division, CERN, CH-1211 Geneva 23, Switzerland
}}\\ 

\vspace*{5mm}
{\it {$^{c}$
Department of Physics, Graduate School of  Science,\\
Osaka University, Toyonaka, Osaka 560-0043, Japan
}}\\

\end{center}

\vspace*{.3cm}
\begin{abstract}
We introduce the strange quark mass
to the Sakai-Sugimoto model of holographic QCD.
We compute mass shifts in the spectra of three-flavor baryons
at the leading order in perturbation in quark masses.
Comparison with experimental data shows an agreement only qualitatively.
\end{abstract}
\vspace{1.5cm}

Oct 2009

\pagestyle{empty}

\end{titlepage}

\setcounter{footnote}{0}
\setcounter{tocdepth}{2}

\newpage

\tableofcontents

\section{Introduction and summary}

Holographic QCD, a string theory realization of QCD in the large $N_c$ limit, 
offers a new approach to QCD.
The Sakai-Sugimoto model \cite{SaSu1,SaSu2} is 
the most successful model of holographic QCD.
Although the original proposal is a holographic dual of {\it massless} QCD,
quark masses can be introduced by 
worldsheet instantons \cite{Hashimoto:2008sr,Aharony:2008an}
(see also \cite{McNees:2008km,Argyres:2008sw,Edalati:2009xc})\footnote{
Another way to introduce the quark masses is 
a tachyon condensation in the D8/$\overline{\mathrm{D8}}$-branes
\cite{Casero:2007ae,Bergman:2007pm,Dhar:2007bz,Dhar:2008um,Jokela:2009tk}.
}.
In the case of two flavors, mass shifts in hadron spectra 
due to the quark masses have been computed in \cite{Hashimoto:2009hj},
where the mass shift of nucleon was found to be consistent with 
lattice QCD results extrapolated to the chiral limit 
\cite{Procura:2003ig,Procura:2006bj,AliKhan:2003cu,Alexandrou:2008tn}.
In the situation that quarks are massless, 
baryons in three-flavor Sakai-Sugimoto model have been studied in \cite{HataMurata}.
However, in {\it real} QCD, 
where we have three flavors below chiral symmetry breaking scale,
mass of the strange quark is heavy compared with up and down quarks. 
Hence it is important to deal with the third flavor as a massive quark, especially to compare with the 
experimental data of hadron spectra.

In this short paper, we introduce the strange quark mass to the Sakai-Sugimoto model.
We compute mass shifts in the spectra of three-flavor baryons, 
at the leading order in the expansion in quark masses.
Our main result is shown in the formula \eqref{3_flavor_gamma_mass_shift},
and the coefficients appearing there are listed in Table~\ref{coefficients_a},
where we used two input parameters $m_\rho$ and $f_\pi$ from experimental data
to fix the parameters of the model, then comparison with experimental data can be meaningful.

Baryons are identified with an instanton-like soliton of the Sakai-Sugimoto model \cite{SaSu1}.
This soliton solution has been studied in detail in \cite{HSSY}
(see also \cite{Hong:2007kx,Hong:2007ay,Hong:2007dq,Park:2008sp,HaSaSu1,HaSaSu2,Kim:2009sr}).
The mass of baryons can be evaluated as a static energy of 
this nontrivial soliton of the gauge theory \cite{HataMurata,HSSY}, 
and also a mass shift of the baryons due to the quark masses can be studied
as an effect of the additional quark mass term in the action \cite{Hashimoto:2009hj}.
Such a kind of mass shift of baryons has been studied also in the context of chiral solitons
(for three-flavor Skyrmions, see e.g.~\cite{Guadagnini:1983uv, Manohar:1984ys}, 
while for a review including other chiral soliton models, see e.g.~\cite{Christov:1995vm}).
However, in the Sakai-Sugimoto model, all the coefficients in the action are fixed 
by only two input parameters from the meson sector.
This is a remarkable difference from many other chiral soliton models\footnote{
The existence of unnecessary modes whose counterparts are absent in QCD, 
is a long-standing problem in holographic QCD.
In this paper, we simply ignore this problem. We expect that a
certain decoupling limit while the QCD scale is fixed can be taken,
once we know systematically the structure of the 
$1/N_c$ and $1/\lambda$ corrections.
}.

In Section~\ref{review_of_2_flavor}, we briefly review
computation of the baryon mass shift in the case of two flavors \cite{Hashimoto:2009hj},
with  giving a brief summary of necessary ingredients in this paper.
We begin with the action of the Sakai-Sugimoto model,
introduce the quark masses, give the baryon configuration,
and after these preparation we compute the baryon mass shift of two-flavor baryons.
Quantum state dependence of $SU(2)$ instanton evaluated here is useful
in Section~\ref{3_flavor_mass_shift}.
We also find that the mass shift of the baryon depends on its spin 
as well as the quantum number of its radial excitation,
which was missing in \cite{Hashimoto:2009hj}.
As a consequence, the mass shift of the delta baryon, which has spin
$3/2$,  
is around 1.5 times larger than that of the proton and the neutron.
This is a characteristic prediction of the Sakai-Sugimoto model, unlike 
conventional chiral soliton models.

In Section~\ref{3_flavor_mass_shift}, we compute 
a shift of the baryon mass due to the quark masses in the case of three flavors.
A soliton solution of the massless three-flavor Sakai-Sugimoto model 
has been studied in \cite{HataMurata}, where a mass formula for baryons was obtained.
The masses of $SU(3)$ baryons are degenerate when quarks are massless.
However, once the $SU(3)$-breaking quark masses are considered, 
there appear mass splittings of baryons.
Using the $SU(3)$ baryon configuration constructed 
by embedding the $SU(2)$ instanton in a $SU(3)$ gauge field,
we obtain a formula of the mass shift of three-flavor baryons,
which depends on the flavor charge,
the radial excitation quantum number, and the spin of the baryons.

Finally, in Section~\ref{discussions},
we discuss the comparison of the mass splittings of baryons
obtained numerically by our computation with those observed in experiments.
In Table~\ref{mass_splitting_hypercharge}, 
we show the mass splittings of baryons with $\Delta Y = 1$ or $\Delta I = 1$,
while we show the mass splittings of baryons with $\Delta I_3 = 1$ 
in Table~\ref{mass_splitting_isospin}.
Agreements of the theoretical and experimental values in these two tables 
are qualitatively good but quantitatively inconclusive.
The reason can be that the physical strange quark mass are no longer very small.
To improve the matching, higher order corrections in the expansion in the strange quark mass are necessary.
In Table~\ref{mass_splitting_8_10}, 
we show a mass difference of an octet baryon and its decuplet counterpart.
This is presented for the study of the known problem 
concerning the overall magnitude of the baryon mass in the Sakai-Sugimoto model.
We observe that, at the leading order in the quark masses,
the inclusion of the quark masses does not improve the situation of this problem.

\section{A review of baryon mass shift}
\label{review_of_2_flavor}

For our computation of the mass shift of the baryons for three flavors
in Section~\ref{3_flavor_mass_shift}, here we give a brief summary of
necessary ingredients, by reviewing \cite{Hashimoto:2009hj}.

\vspace{6mm}
\noindent
\underline{Sakai-Sugimoto model}
\vspace{3mm}

The Sakai-Sugimoto model \cite{SaSu1,SaSu2} is a five-dimensional
gauge theory, in which
Kaluza-Klein(KK)-decomposed gauge fields are mesons, while baryons are 
provided by its solitons.
The action 
is given by 
the following five-dimensional $U(N_f)$ Yang-Mills-Chern-Simons theory 
in a curved background:
\begin{eqnarray}
S=S_\mathrm{YM} + S_\mathrm{CS}, \quad
S_\mathrm{YM}= -\kappa \int d^4 x dz \,\mathrm{tr}
\left[
\frac{1}{2} (1+z^2)^{-\frac{1}{3}} \mathcal{F}_{\mu\nu}^2 + (1+z^2) \mathcal{F}_{\mu z}^2
\right],
\label{SS_action}
\end{eqnarray}
and $S_\mathrm{CS}$ is the 5-form Chern-Simons term.
Here, $\mu, \nu = 0,1,2,3$ are four-dimensional Lorentz indices,
and $z$ is the coordinate of the fifth dimension.
The field strength is defined as 
$\mathcal{F}_{\mu\nu} = \partial_\mu \mathcal{A}_\nu 
- \partial_\nu \mathcal{A}_\mu -i [ \mathcal{A}_\mu, \mathcal{A}_\nu ]$
with the $U(N_f)$ flavor gauge fields $\mathcal{A}_\mu$ and $\mathcal{A}_z$.
They are decomposed as 
\begin{eqnarray}
\mathcal{A} = A + \widehat{A} \frac{\mathbf{1}_{N_f}}{\sqrt{2 N_f}}
= A^a T_a + \widehat{A} \frac{\mathbf{1}_{N_f}}{\sqrt{2 N_f}},
\end{eqnarray}
where $T_a$ ($a=1,\dots, N_f^2-1$) are generators of $SU(N_f)$, 
and $\mathbf{1}_{N_f}$ is the unit matrix of size $N_f$.
We work in the normalization $\mathrm{tr}\, [T_a T_b] = \frac{1}{2}\delta_{ab}$.

This model describes large $N_c$ {\it massless} QCD.
There are two parameters in \eqref{SS_action}: 
a mass scale $M_\mathrm{KK}$ for which we chose a unit 
$M_\mathrm{KK} = 1$, and $\kappa = \lambda N_c/(216\pi^3)$
with the 't~Hooft coupling constant $\lambda$.
The parameter $M_\mathrm{KK}$ can be easily recovered by a dimensional analysis
when we numerically evaluate the mass shift of the baryons.
The two parameters are chosen as 
\begin{eqnarray}
M_\mathrm{KK} = 949 \, \mathrm{[MeV]}, \quad  \kappa = 0.00745
\label{values_of_M_KK_and_kappa}
\end{eqnarray}
to fit the experimental values\footnote{
If one rigorously treats the masslessness,
numerical values in the chiral limit should be used for $f_\pi$ and $m_\rho$.
}
of the $\rho$ meson mass $m_\rho \simeq 776$ MeV 
and the charged pion decay constant $f_\pi \simeq 92.4$ MeV.

\vspace{6mm}
\noindent
\underline{Quark mass in the model} 
\vspace{3mm}


Quark mass can be introduced into the the Sakai-Sugimoto model through worldsheet instantons.
Connecting the D8- and $\overline{\mathrm{D8}}$-branes
corresponds to breaking the chiral symmetry in the model, 
and the worldsheet instanton amplitude induces a quark mass term,
which breaks the chiral symmetry at the Lagrangian level.
This method has been first developed in \cite{Hashimoto:2008sr} and \cite{Aharony:2008an},
but these two papers differ in the regularization of the worldsheet instantons.
In this paper, we follow \cite{Hashimoto:2008sr}.

The relation between the quark mass and the worldsheet instanton is as follows.
Let us connect the D8-branes and the $\overline{\mathrm{D8}}$-branes by D6-branes.
Then we can put a worldsheet instanton whose Euclidean worldsheet has the boundary 
defined by the color D4-branes, the D8-branes, 
the D6-branes and the $\overline{\mathrm{D8}}$-branes.
The worldsheet instanton involves $\bar{q}_L q_R$ vertex 
of the left- and right-handed quarks living at the D8-D4 and the $\overline{\mathrm{D8}}$-D4 intersections, respectively.
This vertex is precisely a quark mass operator.

In the gravity dual picture, 
where two of the corners on the D4-branes 
are smeared out by the background curved geometry,
we can still put the worldsheet instanton in the same manner.
The worldsheet instanton amplitude which includes boundary coupling to the Wilson line 
written in terms of the meson excitation $\mathcal{A}_z$ on the D8-brane,
is given by
\begin{eqnarray}
 \delta S = c \int d^4x \; \mathrm{P\,tr} 
\left[ M \left(
\exp \left[-i \int_{-z_m}^{z_m} \mathcal{A}_z dz \right]
- \mathbf{1}_{N_f} \right) \right] + \mathrm{c.c.} \ ,
\label{worldsheet_boundary_coupling}
\end{eqnarray}
which should correspond to the quark mass term\footnote{
This term breaks a part of the gauge symmetry that the original action \eqref{SS_action} has, 
in a way consistent with the explicit chiral symmetry breaking by
the quark mass term. The reason is as follows.

Let us consider for simplicity the case that
all the $N_f$ quarks share the same mass,
that is, $M$ is proportional to an identity  matrix $\mathbf{1}_{N_f}$.
We have coincident D6-branes in this case.
Because the parts $z=z_m$ and $z=-z_m$ on the D8-brane are connected by
the D6-branes,
the gauge transformations there must be equal.
Thus the gauge symmetry reduces to a restricted part (explicit breaking).
Under this restricted gauge transformation,
\eqref{worldsheet_boundary_coupling} is invariant.
Now, recall that the D8-brane gauge
symmetry at $z \to \pm \infty$, {\it i.e.} at the boundary,
corresponds to the chiral symmetry of the boundary theory
\cite{SaSu1,SaSu2}.
Therefore, if $z_m$ is close to $z=\infty$,
the restriction put on the gauge symmetry by the D6-branes
is consistent with the explicit chiral symmetry breaking by the quark
mass. The breaking pattern is such that
the left and the right chiral symmetries are broken to their
diagonal subgroup.

In general, one can choose different masses for the quarks;
$M$ is not proportional to the identity matrix. The $N_f$ D6-branes
are not on top of each other, but are now at $z=\pm z_m^{(i)}$ where
$i=1,2,\cdots,N_f$.
In this case, the gauge symmetry on the D8-brane is
further restricted, and under this restricted gauge transformation
\eqref{worldsheet_boundary_coupling} is invariant. The remaining gauge symmetry is
consistent with the explicit breaking of the chiral symmetry
($U(N_f)_\mathrm{L} \times U(N_f)_\mathrm{R} \to (U(1))^{N_f}$).
}.
Here  $M$ is a quark mass matrix, and
 $z=z_m$ and $z=-z_m$ specify the location of the D6-brane
on the D8- and the $\overline{\mathrm{D8}}$-branes, respectively.
The constant subtraction $-\mathbf{1}_{N_f}$ makes sure that 
 \eqref{worldsheet_boundary_coupling} vanishes when $\mathcal{A} = 0$.

This term \eqref{worldsheet_boundary_coupling} can be written 
in terms of a pion field $U(x)$. 
The field $\mathcal{A}_z$ relates to the pion field \cite{SaSu1,SaSu2},
\begin{eqnarray}
\mathrm{P} \exp \left[ -i \int_{-\infty}^\infty dz \mathcal{A}_z\right]
= \exp \left[ 2i \pi(x) / f_\pi \right] \equiv U(x) \ ,
\label{A_z_and_pion_field}
\end{eqnarray}
where the decomposition of $\pi(x)$ is defined as $\pi(x) = \pi^a (x) T_a$.
Thanks to this relation, the term \eqref{worldsheet_boundary_coupling} 
can be rewritten as
\begin{eqnarray}
\delta S = \int  d^4x \; \delta L,
\quad
\delta L \equiv c \, \mathrm{tr}
\left[ M(U + U^\dagger-2 \mathbf{1}_{N_f}) \right].
\label{additional_action}
\end{eqnarray}
Here, although $z_m$ is a finite value, we approximate the exponential in 
\eqref{worldsheet_boundary_coupling} by \eqref{A_z_and_pion_field}.

\vspace{6mm}
\noindent
\underline{Meson mass} 
\vspace{3mm}


The definition of $U$ in \eqref{A_z_and_pion_field}
is a standard notation for the pion field in 
chiral perturbation theories,
and it turns out that \eqref{additional_action} reproduces 
a well-known form of the quark mass term 
in the chiral perturbation theories \cite{Gasser:1983yg,Gasser:1984gg}.
A constant $c$ appears as an overall factor in \eqref{additional_action},
but it is difficult to evaluate the constant $c$ accurately 
because it comes from sub-leading contributions of the worldsheet instantons in the curved space.
As we shall see, using meson masses can be an alternative to the explicit use of 
the constant $c$ and the quark masses.
We show the cases $N_f=2$ and $N_f=3$ in turn.

In the case that $N_f=2$,
\eqref{additional_action} can be parametrized by a pion mass $m_\pi$.
The quark mass matrix is given by $M = \mathrm{diag} (m_u, m_d)$, 
where $m_u$ and $m_d$ are the up and down quark masses, respectively.
They are related with the pion mass as
\begin{eqnarray}
m_{\pi}^2 = \frac{2c}{f_\pi^2} (m_u+m_d).
\label{2_flavor_quark_mass_meson_mass}
\end{eqnarray}
Note that charged and neutral pions have the same masses in this case, 
at the leading order.

In the case that $N_f = 3$, we need three meson masses.
The quark mass matrix is given by $M = \mathrm{diag} (m_u, m_d, m_s)$, 
where $m_s$ is the strange quark mass.
Using the charged pion mass $m_{\pi^\pm}$, the charged kaon mass $m_{K^\pm}$ and 
the neutral kaon mass $m_{K^0,\bar{K}^0}$, we have
\begin{eqnarray}
m_{\pi^\pm}^2 = \frac{2c}{f_\pi^2} (m_u+m_d),\quad 
m_{K^\pm}^2 = \frac{2c}{f_\pi^2} (m_u+m_s),\quad 
m_{K^0,\bar{K}^0}^2 = \frac{2c}{f_\pi^2} (m_d+m_s).
\label{3_flavor_quark_mass_meson_mass}
\end{eqnarray}
Since these are the simplest relations between quark and meson masses, 
we use these as input parameters.

In the rest of this section, we work with $N_f=2$.

\vspace{6mm}
\noindent
\underline{Baryon in the model} 
\vspace{3mm}

The baryon is identified as an instanton soliton localized in the
four-dimensional $x^M$ space ($M=1,2,3,z$) \cite{SaSu1}.
The instanton number of the Yang-Mills theory \eqref{SS_action} 
corresponds to the baryon number. 
An explicit solution of the equations of motion of the action \eqref{SS_action} 
has been obtained in \cite{HSSY}.
Here we collect the part we need.

The part relevant to our computation is the $\mathcal{A}_z$ component of the solution.
The non-Abelian part is identical to the BPST instanton,
which, in a singular gauge, is given by
\begin{eqnarray}
A_z = \left( \frac1{\xi^2} - \frac1{\xi^2 +\rho^2} \right) 
(x^i-X^i) \tau_i \equiv A_{\mathrm{BPST}}\ ,
\label{singular_gauge_BPST_instanton}
\end{eqnarray}
while the $U(1)$ part is simply $\widehat{A}_z = 0$. 
Here 
$\xi \equiv \sqrt{(z-Z)^2 + |\vec{x}-\vec{X}|^2}$, and
$\tau_i$ ($i = 1,2,3$) are the Pauli matrices.
In \eqref{singular_gauge_BPST_instanton}, 
five of the eight moduli parameters of the instanton come in explicitly:
$\rho$ is the size of the instanton, and 
$X^M = (X^1,X^2,X^3,Z) = (\vec{X},Z)$
is the location of the instanton.
As well as these five, there are three $SU(2)$ rotation moduli.

The classical value of the moduli is also obtained in \cite{HSSY} as\footnote{
The action \eqref{SS_action} is at the leading order in $1/\lambda$ expansion,
while the size of the instanton \eqref{HSSY_classical_rho_z} is $\mathcal{O}(\lambda^{-1/2})$.
In order to take sub-leading corrections into account, 
the authors of  \cite{HSSY} have studied also the Dirac-Born-Infeld(DBI) action,
and they obtained the same classical value of the moduli as in \eqref{HSSY_classical_rho_z}.
However, the DBI corrections do not suffice all the $1/\lambda$ corrections.
In this paper, we simply assume \eqref{SS_action} as our starting point.
}
\begin{eqnarray}
\rho_\mathrm{cl}^2
= \frac{27\pi}{\lambda} \sqrt{\frac{6}{5}} \ ,
\quad
Z_\mathrm{cl}=0 \ .
\label{HSSY_classical_rho_z}
\end{eqnarray}
We can put $X^i = 0$ without loss of generality.

\vspace{6mm}
\noindent
\underline{Mass shift of the baryons} 
\vspace{3mm}

Now we compute the baryon mass shift, 
including the dependence on the quantum states of the baryon.
At the leading order in the quark mass, 
a classical shift of the baryon mass is simply given by
\begin{eqnarray}
\delta M = -\int d^3x\;\delta L[A^\mathrm{cl}],
\label{mass_shift_lagrangian}
\end{eqnarray}
where $A^\mathrm{cl}$ is the classical solution \eqref{singular_gauge_BPST_instanton}.
It receives corrections when we consider the quantum states of the baryon.

We begin with the $z$ integral in \eqref{A_z_and_pion_field}.
As the $A_z$ configuration \eqref{singular_gauge_BPST_instanton}
is proportional to a matrix $x^i \tau_i$ for any value of $z$, 
 the path ordering in \eqref{A_z_and_pion_field}
reduces to an Abelian problem \cite{Atiyah:1989dq}.
Hence we obtain
\begin{eqnarray}
U = \exp \left[ i f(r) \hat{x}^i \tau_i \right]
\label{SU2_hedgehog_soliton}
\end{eqnarray}
with
\begin{eqnarray}
f(r)= \pi \left[ 1-\frac{1}{\sqrt{1+\rho^2/r^2}} \right],
\quad
r \equiv \sqrt{(x^1)^2 + (x^2)^2 + (x^3)^2},
\quad
\hat{x}^i \equiv \frac{x^i}{r} \ .
\label{f_after_z_integration}
\end{eqnarray}
Further, using \eqref{SU2_hedgehog_soliton} and the relation \eqref{2_flavor_quark_mass_meson_mass},
we obtain a formula for the mass shift,
\begin{eqnarray}
\delta M=  4 \pi f_\pi^2 \rho^3 m_\pi^2 \times 1.104 \ . 
\label{2_flavor_classical_baryon_mass_shift}
\end{eqnarray}

Let us consider the quantum state dependence of the baryon.
Each of the baryon state can be specified by the quantum numbers
$\{ I(=J), I_3, n_\rho, n_Z \}$,
where $I$ is the isospin, $J$ is the spin,
and $n_\rho$ and $n_z$ are quantum numbers associated 
with the moduli $\rho$ and $Z$, respectively. 
However, the $Z$-dependence disappears because 
it comes in as $z-Z$ in the baryon solution \eqref{singular_gauge_BPST_instanton},
and disappears in \eqref{2_flavor_classical_baryon_mass_shift}.
Dependence on the $SU(2)$ rotation moduli also disappears 
because $U + U^\dagger$ is proportional to the unit matrix,
and the $SU(2)$ rotation $U + U^\dagger \to G (U + U^\dagger) G^\dagger$ gives no effect.
Therefore, there is no dependence on $n_Z$ and $I_3$, once $I$ is specified.
The mass of proton and neutron are equal here.

The mass shift, however, depends on the modulus $\rho$.
We can show that the expectation value of $\rho^n$
is labeled by $n_\rho$ and the spin $l/2 (=J)$; 
we write this as $\langle \rho^n \rangle_{n_\rho, l}$.
To compute this, we use
the eigenfunction for the quantization of $\rho$ obtained in \cite{HSSY},
\begin{eqnarray}
R(\rho\, ; n_\rho, l) 
= C_{n_\rho,l} \, e^{- c_\rho \rho^2 /2} \, \rho^{\beta_l -2} \, 
F(-n_\rho,\beta_l; c_\rho \rho^2),
\label{R_rho_eigenfunction}
\end{eqnarray}
where $\beta_l \equiv 1+ \sqrt{(l+1)^2 + 4 N_c^2/5}$, $c_\rho \equiv 16 \pi^2 \kappa / \sqrt{6}$,
and $F(\alpha,\gamma;z)$ is the confluent hypergeometric function defined by
$
F(\alpha,\gamma;z) 
\equiv \sum_{k=0}^\infty \frac{(\alpha)_k}{(\gamma)_k} \frac{z^k}{k!}
$
with $(\alpha)_k\equiv \alpha(\alpha+1)\cdots(\alpha+k-1)$.
Here $C_{n_\rho,l}$ is the normalization factor, 
and we can normalize $R(\rho\, ; n_\rho, l)$ as
\begin{eqnarray}
\int_0^\infty d\rho \; \rho^3 R(\rho \, ; n_\rho, l)^2 = 1,
\end{eqnarray}
for each $n_\rho$ and $l$. 
Let us consider the case that $n_\rho =0$.
Then, $C_{0,l}$ is computed as
\begin{eqnarray}
C_{0,l} =
\sqrt{
\frac{2 \, c_\rho^{\beta_l}}{\Gamma(\beta_l)}
}.
\end{eqnarray}
Using this, 
we can compute the expectation value of $\rho^n$ as
\begin{eqnarray}
\langle \rho^n \rangle_{n_\rho = 0, l}  =
\int_0^\infty d\rho \; \rho^{3+n} R(\rho \, ; 0, l)^2  
= \frac{\Gamma(\beta_l + n/2)}{c_\rho^{n/2} \Gamma(\beta_l)}.
\label{expectation_value_rhon_when_nrho0}
\end{eqnarray}
It is straightforward to repeat this computation for other values 
of $n_\rho$.
For example, for $n_\rho=1$, we have
\begin{eqnarray}
\langle \rho^n \rangle_{n_\rho = 1, l} 
= \left( 1 + \frac{n(n+2)}{4\beta_l} \right)
 \frac{\Gamma(\beta_l + n/2)}{c_\rho^{n/2} \Gamma(\beta_l)} 
 = \left( 1 + \frac{n(n+2)}{4\beta_l} \right) \langle \rho^n \rangle_{n_\rho = 0, l}.
 \label{expectation_value_rhon_when_nrho1}
\end{eqnarray}

To obtain the baryon mass shift for a given quantum state of the baryon,
we use the formula \eqref{expectation_value_rhon_when_nrho0}
with $n=3$ and rewrite $\kappa$ in terms of $f_\pi$ as 
$\kappa = \pi f_\pi^2/4$,  
\begin{eqnarray}
\langle \rho^3 \rangle_{n_\rho = 0, l} 
= \frac{1}{f_\pi^3} \left( \frac{\sqrt{6}}{4\pi^3} \right)^{3/2}
\frac{\Gamma(\beta_l +3/2)}{\Gamma(\beta_l)}.
\label{n_rho_0_ratio}
\end{eqnarray}
Using \eqref{n_rho_0_ratio}, 
we obtain the mass shift for the 
baryon with $n_\rho = 0$ as
\begin{eqnarray}
\frac{\delta M_{n_\rho=0,l }}{m_\pi^2} = 
1.104 \times \frac{6^{3/4}}{2 \, \pi^{7/2}f_\pi} \frac{\Gamma(\beta_l  +3/2)}{\Gamma(\beta_l)}.
\label{two_flavor_mass_shift_formula}
\end{eqnarray}
Although two parameters $f_\pi$ and $m_\rho$ can be fit by \eqref{values_of_M_KK_and_kappa}, 
only $f_\pi$ is relevant for evaluating the baryon mass shift.
If we use $f_\pi = 92.4$ MeV and $N_c = 3$, \eqref{two_flavor_mass_shift_formula} becomes
\begin{eqnarray}
&& \hspace{-3ex}  \frac{\delta M_{n_\rho=0, l=1}}{m_\pi^2}
= 4.1 \;\mathrm{[GeV^{-1}]} , 
\label{two_flavor_41_shift}
\\
&& \hspace{-3ex}  \frac{\delta M_{n_\rho=0, l=3}}{m_\pi^2}
= 6.2 \;\mathrm{[GeV^{-1}]} ,
\label{two_flavor_62_shift}
\end{eqnarray}
where the first line corresponds to the mass shift of the proton 
and the neutron,
which share the same mass shift in the two-flavor case, 
and the second line corresponds to that of the delta baryon.
The worldsheet instantons in Sakai-Sugimoto model predict 
that the mass shift of the delta baryon is around 1.5 times larger
than that of the nucleon and the proton, at the leading order in expansion in pion mass.
The result \eqref{two_flavor_41_shift} is consistent with values obtained in chiral extrapolation of
results of lattice QCD
\cite{Procura:2003ig,Procura:2006bj,AliKhan:2003cu,Alexandrou:2008tn}.

\section{Three flavors}
\label{3_flavor_mass_shift}

Now we consider dependence on three-flavor quark masses
by introducing the strange quark mass.
When quarks are massless, the baryon solution in three-flavor Sakai-Sugimoto model 
has been studied in \cite{HataMurata}.
It was found there that 
the classical moduli are given in this case again by \eqref{HSSY_classical_rho_z}
(we can put $X^i = 0$).
As a correction to the massless case,
we study the additional action given by \eqref{additional_action} with $N_f = 3$
to evaluate a shift of the baryon mass.

Here we pay attention to how to construct the $SU(3)$ baryon states, 
for we can compute other than that in a way similar to Section~\ref{review_of_2_flavor}.
The $SU(3)$ baryon states, whose wave functions are written by 
the seven $SU(3)$ rotation moduli of the twelve $SU(3)$ instanton moduli, 
can be obtained in the following three steps;
we first embed the $SU(2)$ classical soliton solution into the $SU(3)$ gauge field,
next consider the $SU(3)$ rotation moduli, and finally project this rotated-soliton configuration 
onto each baryon state.
For this, instantons can allow the techniques developed for Skyrmions
\cite{Guadagnini:1983uv}
(see also \cite{Mazur:1984yf,SriRam:1984xt,Praszalowicz:1985bt,Chemtob:1985ar,Yabu:1987hm}).

The $SU(2)$ BPST instanton solution \eqref{singular_gauge_BPST_instanton}
can be embedded in the $SU(3)$ gauge field as
\begin{eqnarray}
A_z =
\begin{pmatrix}
A_{\mathrm{BPST}} & 0 \\ 0 & 0
\end{pmatrix}.
\label{embedding_su2_to_su3}
\end{eqnarray}
This $A_z$ configuration is still proportional to a matrix $x^i \tau_i$ 
as in the case of two flavors. 
Therefore the path ordering in \eqref{A_z_and_pion_field} 
can be evaluated in the same way as in Section~\ref{review_of_2_flavor}.
As a result, we obtain
\begin{eqnarray}
U_0(x)=
\begin{pmatrix}
\exp[i f(r) \hat{x}^i \tau_i] & 0 \\ 0 & 1
\end{pmatrix},
\label{su3_hedgehog}
\end{eqnarray}
where $f(r)$ appearing here is the same as that appearing in \eqref{f_after_z_integration}.
Here $U_0$ represents the classical embedding of the $SU(2)$ instanton.

The classical embedding \eqref{su3_hedgehog} should be rotated so as to include
the $SU(3)$ baryon states.
Note that although the actual moduli space is $SU(3)/U(1)$,
one can simply consider rotations in $SU(3)$ space, with imposing 
a constraint on harmonic functions on $SU(3)$ \cite{Witten:1983tx}.
Hence we can simply put the $SU(3)$-rotated form
\begin{eqnarray}
U(x)=G U_0(x) G^\dagger, \quad G \in SU(3),
\end{eqnarray}
in \eqref{additional_action},
and we  obtain \cite{Guadagnini:1983uv}
\begin{eqnarray}
\delta L 
&=& -\frac{4c}{3}(1-\cos f(r))
\Bigg[
(m_u+m_d+m_s) \nonumber \\
&& \left. 
- \frac{\sqrt{3}}{2}(m_d-m_u)D^{(8)}_{38}(G) 
- \frac{2m_s-m_u-m_d}{2}D^{(8)}_{88}(G) \right],
\label{mass_term_written_by_D_func}
\end{eqnarray}
where
\begin{eqnarray}
D^{(8)}_{ab}(G) = \frac{1}{2} \mathrm{tr}(G^\dagger \lambda_a G \lambda_b)
\end{eqnarray}
is the Wigner's D function for the adjoint representation of $SU(3)$,
and $\lambda_a$ ($a = 1,\dots, 8$) are the Gell-Mann matrices.
Further, we use \eqref{3_flavor_quark_mass_meson_mass} 
to write the mass shift as a function of meson masses. 
Then we obtain
\begin{eqnarray}
\delta M 
&=& 4 \pi f_\pi^2 \rho^3 \times 1.104 \times \frac{1}{3}
\left[ \left(1 - \sqrt{3} D^{(8)}_{38}(G) - D^{(8)}_{88}(G) \right) m_{K^0,\bar{K}^0}^2 \right. \nonumber \\
&& \left. + \left(1 + \sqrt{3} D^{(8)}_{38}(G) - D^{(8)}_{88}(G) \right) m_{K^\pm}^2
+\left(1 + 2 D^{(8)}_{88}(G)\right) m_{\pi^\pm}^2 \right].
\end{eqnarray}
This mass shift depends on the quantum states of $SU(3)$ rotation and $\rho$ moduli,
while it does not depend on $Z$.

Now we project the operators $D^{(8)}_{38}(G) $ and $D^{(8)}_{88}(G)$
onto each baryon state.
The wave function of the $SU(3)$ baryons is given by \cite{Guadagnini:1983uv, Manohar:1984ys}
\begin{eqnarray}
\Psi_B(G)
= \sqrt{\mathrm{dim} \; r} (-1)^{J_3 + 1/2} D^{(r)}_{Y, I, I_3;1, J, -J_3}(G)^\ast
\equiv \sqrt{\mathrm{dim} \; r} (-1)^{J_3 + 1/2} D^{(r)}_{\mu \nu}(G)^\ast.
\label{baryon_wave_function}
\end{eqnarray}
Here $r$ is a label for the representation of $SU(3)$; 
$Y, I, I_3$ are the quantum numbers of $SU(3)_{\mathrm{flavor}}$,
where $I$ is the isospin, $I_3$ is its third component, and $Y$ is the hypercharge;
$J, J_3$ are those of $SU(2)_{\mathrm{spin}} \subset SU(3)_\mathrm{right}$,
where $J$ is the spin and $J_3$ is its third component.
A constraint $Y_R = N_c / 3 =1$ (for $N_c =3$) is imposed\footnote{
This is expected to come from the Chern-Simons term 
of the Sakai-Sugimoto model. A problem on the derivation of the
constraint has been reported \cite{HataMurata},
and here we simply assume this constraint.
}
on the hypercharge of $SU(3)_\mathrm{right}$.
In the last expression of \eqref{baryon_wave_function},
 $\mu$ and $\nu$ represent $\{ I, \, I_3, \, Y \}$ and $\{ J, \, -J_3, \, 1 \}$
in abbreviated form, respectively.
Using \eqref{baryon_wave_function},
the expectation value of the D functions can be evaluated as
\begin{eqnarray}
\langle D^{(8)}_{ab}(G) \rangle_B 
&=& \int dG \; \Psi_{B}^\ast (G) D^{(8)}_{ab}(G) \Psi_{B}(G) \nonumber \\
&=& \sum_\gamma
\begin{pmatrix}
8 & r & r_{\gamma} \\ a & \mu & \mu
\end{pmatrix}
\begin{pmatrix}
8 & r & r_{\gamma} \\ b & \nu & \nu
\end{pmatrix}.
\end{eqnarray}
The summation is taken over all occurrences of the representation 
$r$ in the product of the representation $8$ and the other $r$.
A table of the $SU(3)$ Clebsch-Gordan coefficients \cite{deSwart:1963gc} 
is given, for example, in \cite{McNamee:1964xq}.
The results are shown in Table~\ref{sandwiched_D_functions} \cite{Guadagnini:1983uv}.

\begin{table}[htdp]
\begin{center}
\caption{$\langle D^{(8)}_{38}(G)/I_3 \rangle_B$ and $\langle D^{(8)}_{88}(G) \rangle_B$}
\label{sandwiched_D_functions}
\renewcommand{\arraystretch}{1.3}
\begin{tabular}{ccc}
\hline
{\bf 8} & $\langle D^{(8)}_{38}(G)/I_3 \rangle_B$ & $\langle D^{(8)}_{88}(G) \rangle_B$ \\ \hline
$N$ & $\frac{1}{5\sqrt{3}}$ & $\frac{3}{10}$ \\
$\Lambda$ & $0$ & $\frac{1}{10}$ \\
$\Sigma$ & $\frac{1}{2\sqrt{3}}$ & $-\frac{1}{10}$ \\
$\Xi$ & $\frac{4}{5\sqrt{3}}$ & $-\frac{1}{5}$ \\ \hline
\end{tabular}
\quad
\begin{tabular}{ccc}
\hline
{\bf 10} & $\langle D^{(8)}_{38}(G)/I_3 \rangle_B$ & $\langle D^{(8)}_{88}(G) \rangle_B$ \\ \hline
$\Delta$ & $\frac{1}{4\sqrt{3}}$ & $\frac{1}{8}$ \\
$\Sigma^\ast$ & $\frac{1}{4\sqrt{3}}$ & $0$ \\
$\Xi^\ast$ & $\frac{1}{4\sqrt{3}}$ &  $-\frac{1}{8}$\\
$\Omega$ & $0$ & $-\frac{1}{4}$ \\ \hline
\end{tabular}
\end{center}
\end{table}%

Using Table~\ref{sandwiched_D_functions} to take into account the $SU(3)$ baryon state dependence,
we obtain a formula for the mass shifts in the baryon spectra,
at the leading order in the quark masses which are rewritten in the meson mass squared,
\begin{eqnarray}
\delta M_B = 4 \pi f_{\pi}^2 \rho^3 \times 1.104 \times \frac{1}{3}
\left( a_0 m_{K^0,\bar{K}^0}^2 + a_K m_{K^\pm}^2 + a_\pi m_{\pi^\pm}^2 \right) \ ,
\label{formula_for_three_flavor_baryon_mass_shift}
\end{eqnarray}
with the coefficients $a_0, \, a_K$ and $a_\pi$ listed in Table~\ref{coefficients_a}.
This result \eqref{formula_for_three_flavor_baryon_mass_shift} is symmetric under 
the interchange of two flavors:
$u \leftrightarrow d$, $u \leftrightarrow s$, and $d \leftrightarrow s$.
This symmetry is manifest in Table~\ref{coefficients_a}.

\begin{table}[htdp]
\begin{center}
\caption{The values of $a_0, \, a_K$ and $a_\pi$ for each baryon state}
\label{coefficients_a}
\renewcommand{\arraystretch}{1.3}
\begin{tabular}{ccccccccc}
\hline
{\bf 8} & $P$ & $N$ & $\Lambda$ & $\Sigma^+$ & 
$\Sigma^0$ & $\Sigma^-$ & $\Xi^0$ & $\Xi^-$ \\ \hline
$a_0$ & $\frac{3}{5}$ & $\frac{4}{5}$ & $\frac{9}{10}$ & $\frac{3}{5}$ & 
$\frac{11}{10}$ & $\frac{8}{5}$ & $\frac{4}{5}$ & $\frac{8}{5}$ \\
$a_K$ & $\frac{4}{5}$ & $\frac{3}{5}$ & $\frac{9}{10}$ & $\frac{8}{5}$ & 
$\frac{11}{10}$ & $\frac{3}{5}$ & $\frac{8}{5}$ & $\frac{4}{5}$ \\
$a_\pi$ & $\frac{8}{5}$ & $\frac{8}{5}$ & $\frac{6}{5}$ & $\frac{4}{5}$ & 
$\frac{4}{5}$ & $\frac{4}{5}$ & $\frac{3}{5}$ & $\frac{3}{5}$ \\ \hline
\end{tabular}
\\ \vspace*{5pt}
\begin{tabular}{ccccccccccc}
\hline
{\bf 10} & $\Delta^{++}$ & $\Delta^+$ & $\Delta^0$ & $\Delta^-$ & $\Sigma^{\ast +}$ & 
$\Sigma^{\ast 0}$ & $\Sigma^{\ast -}$ & $\Xi^{\ast 0}$ & $\Xi^{\ast -}$ & $\Omega^-$ \\ \hline
$a_0$ & $\frac{1}{2}$ & $\frac{3}{4}$ & $1$ & $\frac{5}{4}$ & $\frac{3}{4}$ & 
$1$ & $\frac{5}{4}$ & $1$ & $\frac{5}{4}$ & $\frac{5}{4}$ \\
$a_K$ & $\frac{5}{4}$ & $1$ & $\frac{3}{4}$ & $\frac{1}{2}$ & $\frac{5}{4}$ & 
$1$ & $\frac{3}{4}$ & $\frac{5}{4}$ & $1$ & $\frac{5}{4}$ \\
$a_\pi$ & $\frac{5}{4}$ & $\frac{5}{4}$ & $\frac{5}{4}$ & $\frac{5}{4}$ & $1$ & 
$1$ & $1$ & $\frac{3}{4}$ & $\frac{3}{4}$ & $\frac{1}{2}$ \\ \hline
\end{tabular}
\end{center}
\end{table}%

We evaluate the expectation value of $\rho^3$ 
in \eqref{formula_for_three_flavor_baryon_mass_shift}.
We can evaluate its expectation value 
by using the eigenfunction obtained in the case of three flavors \cite{HataMurata},
\begin{eqnarray}
R(\rho; \, n_\rho,(p,q),l) = 
C_{n_\rho(p,q)l}  \,e^{- c_\rho \rho^2 /2} \, \rho^{\beta_{(p,q)l} -(\eta+1)/2} F(-n_\rho, \beta_{(p,q)l}\, ; c_\rho \rho^2)  ,
\label{rho_eigenfunction_three_flavors}
\end{eqnarray}
where $\eta = 8$ is the dimension of $SU(3)$, $C_{n_\rho(p,q)l} $ is a normalization factor, and
\begin{eqnarray}
\beta_{(p,q)l} \equiv 
1+\sqrt{\frac{(\eta-1)^2}{4} +  \frac{2 N_c^2}{15} + \frac{8}{3} \left( p^2 + q^2 + 3(p+q) + pq \right) - l(l+2)}.
\label{beta_pql_three_flavors}
\end{eqnarray}
Here the $(p,q) = (1,1), l=1$ and the $(p,q) = (3,0), l=3$ cases correspond to 
the octet and the decuplet, respectively.
The eigenfunction is normalized as $\int d\rho \rho^{\eta} R^2 =1$.
Starting with \eqref{rho_eigenfunction_three_flavors}, 
we obtain the expectation value of $\rho^n$ in the $n_\rho=0$ case as
\begin{eqnarray}
\langle \rho^n \rangle_{n_\rho = 0,(p,q),l} 
= \frac{\Gamma(\beta_{(p,q)l} + n/2)}{c_\rho^{n/2} \Gamma(\beta_{(p,q)l})}.
\label{rhon_expectation_value_three_flavors}
\end{eqnarray}
We find that it is necessary to replace $\beta_l$ of the two-flavor case
with $\beta_{(p,q)l}$ of the three-flavor case.
Computations for $n_\rho \neq 0$ cases can be done in similar manners.

Thus we obtain the mass shift for the baryon state with $n_\rho = 0$ as
\begin{eqnarray}
\delta M_{B, n_\rho = 0,(p,q),l} = 
&& \hspace{-3ex} 
1.104 \times \frac{6^{3/4}}{2 \, \pi^{7/2}f_\pi} \frac{\Gamma(\beta_ {(p,q)l} +3/2)}{\Gamma(\beta_ {(p,q)l})} 
\nonumber \\
&& \hspace{-3ex} 
\times \frac{1}{3}
\left( a_0 m_{K^0}^2 + a_K m_{K^\pm}^2 + a_\pi m_{\pi^\pm}^2 \right),
\label{3_flavor_gamma_mass_shift}
\end{eqnarray}
where $\beta_ {(p,q)l}$ is defined in \eqref{beta_pql_three_flavors},
and the coefficients $a_0, \, a_K$ and $a_\pi$ are listed in Table~\ref{coefficients_a}.
This is our main result for the mass shift 
at the leading order in the quark masses from the chiral limit.
Using $f_\pi = 92.4$ MeV and $N_c=3$, we obtain
\begin{eqnarray}
\delta M_{B, n_\rho = 0, \mathrm{oct}} =
\frac{1}{3}
\left( a_0 m_{K^0}^2 + a_K m_{K^\pm}^2 + a_\pi m_{\pi^\pm}^2 \right)
\times 7.9 \; \mathrm{[GeV^{-1}]} ,
\nonumber \\
\delta M_{B, n_\rho = 0,\mathrm{dec}} =
\frac{1}{3}
\left( a_0 m_{K^0}^2 + a_K m_{K^\pm}^2 + a_\pi m_{\pi^\pm}^2 \right)
\times 9.5 \; \mathrm{[GeV^{-1}]},
\label{3_flavor_41_mass_shift}
\end{eqnarray}
for octet (spin 1/2) and decuplet (spin 3/2) baryons, respectively.
In Section~\ref{discussions}, we discuss a comparison with experimental data.

We can evaluate the $\rho$-dependence also for the $n_\rho = 1$ excitation, 
which includes the Roper excitation.
We have the following ratio of the $n_\rho=1$ and the $n_\rho=0$ cases:
$\langle \rho^3 \rangle_{n_\rho = 1, (p,q)l} /\langle \rho^3 \rangle_{n_\rho = 0,(p,q)l}
= (\beta_ {(p,q)l} + 15/4)/\beta_ {(p,q)l}$.
Hence the mass shifts for the $n_\rho = 1$ excited baryon states
are given by multiplying the right-hand side of \eqref{3_flavor_gamma_mass_shift}
by $(\beta_ {(p,q)l} + 15/4)/\beta_ {(p,q)l}$.
As a consequence,
$7.9$ and $9.5$ in \eqref{3_flavor_41_mass_shift} are replaced with $12$ and $14$, respectively.

Lattice QCD simulation with a light strange quark has not been carried
out yet. We hope that lattice computations of QCD in the future,
preferably with large $N_c$, may reproduce our analytic computations
\eqref{3_flavor_gamma_mass_shift} and \eqref{3_flavor_41_mass_shift}.

A comment on higher order corrections is in order.
Our result \eqref{3_flavor_gamma_mass_shift} is at the leading order in the quark masses 
breaking the flavor symmetry, and also in the $1/N_c$ expansion.
There are higher order corrections to the action of the Sakai-Sugimoto model.
The higher order corrections in the $1/N_c$ expansion are from the higher order string loops,
while those in the expansion in the quark masses can be obtained 
by the worldsheet instanton amplitudes with higher instanton numbers \cite{Hashimoto:2008sr}.
It is important to compute such higher order corrections,
to get closer to real QCD in the holographic approach.

\section{Discussions}
\label{discussions}
We compare the mass splittings of baryons numerically evaluated from \eqref{3_flavor_41_mass_shift}
with the observed baryon spectra in PDG \cite{Amsler:2008zzb}.
Note that the mass shift \eqref{3_flavor_41_mass_shift} is obtained in perturbation 
at the leading order in quark mass from the chiral limit,
and one should not expect our theoretical spectra to agree well, 
particularly with observed values for baryons containing strangeness.
The comparison here is a first step to include contributions from masses of three-flavor quarks.

We use the following experimental values as input parameters of meson masses:
\begin{eqnarray}
m_{\pi^\pm} = 140 \; \mathrm{[MeV]}, \quad
m_{K^\pm} = 494 \; \mathrm{[MeV]}, \quad
m_{K^0,\bar{K}^0} = 498  \; \mathrm{[MeV]}.
\end{eqnarray}
Together with \eqref{values_of_M_KK_and_kappa}, we use five input parameters, 
all of which come from the meson sector.

\begin{table}[htdp]
\begin{center}
\caption{Mass splitting of baryons with $\Delta Y = 1$ or $\Delta I = 1$}
\label{mass_splitting_hypercharge}
\vspace*{1mm}
\begin{tabular}{cccccc} \hline
$\mathbf{8}$ & $\Delta m_{\Lambda - N}$ 
& $\Delta m_{\Sigma^0 - N}$ & $\Delta m_{\Sigma^0 - \Lambda}$ 
& $\Delta m_{\Xi^0 - \Lambda}$ & $\Delta m_{\Xi^0 - \Sigma^0}$ \\ \hline
theory [MeV] & $2.4 \times 10^2$ & $4.8 \times 10^2$ & $2.4 \times 10^2$ &  $3.5 \times 10^2$ & $1.2 \times 10^2$ \\
experiment [MeV] & $1.8 \times 10^2$  & $2.5 \times 10^2$  & 77 & $2.0 \times 10^2$  & $1.2 \times 10^2$  \\ \hline
\end{tabular}
\\ \vspace*{5pt}
\begin{tabular}{ccccccccc} \hline
$\mathbf{10}$ & $\Delta m_{\Sigma^{\ast 0} - \Delta^0}$ &
$\Delta m_{\Xi^{\ast 0} - \Sigma^{\ast 0}}$ & $\Delta m_{\Omega^- - \Xi^{\ast 0}}$ \\ \hline
theory [MeV] &  $1.8 \times 10^2$ & $1.8 \times 10^2$ & $1.8 \times 10^2$ \\
experiment [MeV] & $1.5 \times 10^2$  & $1.5 \times 10^2$  & $1.4 \times 10^2$  \\ \hline
\end{tabular}
\end{center}
\end{table}%

The  mass splittings of baryons with $\Delta Y = 1$ or $\Delta I = 1$ 
are shown in Table~\ref{mass_splitting_hypercharge},
where the mass difference of two baryons  $B_1$ and $B_2$ is written as
\begin{eqnarray}
\Delta m_{B_1 - B_2} \equiv m_{B_1} - m_{B_2}.
\end{eqnarray}
The entry $\Delta m_{\Sigma^0 - \Lambda}$ corresponds to $\Delta I = 1$, 
while other entries are for baryons with $\Delta Y = 1$.
Here we adopt neutral baryons except for the $\Omega^-$ baryon
in order to avoid the effects of the electromagnetic interactions,
which is not considered here.

This result, at the leading order in quark masses,
qualitatively captures a tendency of the baryon mass splittings,
while quantitatively it is unsatisfactory.
This could be because theoretical values are obtained by a linear extrapolation 
to the physical strange quark mass ($70$ MeV to $130$ MeV), 
which is no longer very small.
To improve, we need higher order corrections, 
especially in the expansion in the strange quark mass.

\begin{table}[htdp]
\begin{center}
\caption{Mass splitting between baryons with $\Delta I_3 = 1$}
\label{mass_splitting_isospin}
\vspace*{1mm}
\begin{tabular}{ccccc} \hline
$\mathbf{8}$ & $\Delta m_{N - P}$ & $\Delta m_{\Sigma^- - \Sigma^0}$ &
$\Delta m_{\Sigma^0 - \Sigma^+}$ & $\Delta m_{\Xi^- - \Xi^0}$ \\ \hline
theory [MeV] & 2.1 & 5.1 & 5.1 & 8.2\\
experiment [MeV] & 1.3 & 4.8 $\pm$ 0.1& 3.3 $\pm$ 0.1 & 6.9 $\pm$ 0.3 \\ \hline
\end{tabular}
\\ \vspace*{5pt}
\begin{tabular}{ccccc} \hline
$\mathbf{10}$ & $\Delta m _{\Delta \mathrm{baryons}}$ & 
$\Delta m_{\Sigma^{\ast -} - \Sigma^{\ast 0}}$ &
$\Delta m_{\Sigma^{\ast 0} - \Sigma^{\ast +}}$ & $\Delta m_{\Xi^{\ast -} - \Xi^{\ast 0}}$ \\ \hline
theory [MeV] & 3.1 & 3.1 & 3.1 & 3.1\\
experiment [MeV] & $(\lesssim 2)$& 3.5 $\pm$ 1.5 & 0.9 $\pm$ 1.4 & 3.2 $\pm$ 0.9\\ \hline
\end{tabular}
\end{center}
\end{table}%

The mass splittings of baryons in an isospin multiplet and with $\Delta I_3=1$ 
are shown in Table~\ref{mass_splitting_isospin}.
We again find a qualitatively good but quantitatively inconclusive agreement
in the comparison with experiments.
One of the reasons is the same as in the previous paragraph; 
although the $\Delta I_3=1$ mass splittings are proportional to a small value $m_d - m_u$,
which is translated into the difference between $m_{K^\pm}$ and $m_{K^0,\bar{K}^0}$,
the computation here is at the leading order in the strange quark mass.
In addition to this, we expect that the electromagnetic interactions 
ignored here can be effective for the $\Delta I_3=1$ mass splittings.

\begin{table}[htdp]
\begin{center}
\caption{Mass difference of an octet baryon and its decuplet counterpart}
\label{mass_splitting_8_10}
\vspace*{1mm}
\begin{tabular}{ccccc} \hline
$B^\ast - B$ & $\mathbf{10}-\mathbf{8}$ & $\Delta^{0}-N$ & 
$\Sigma^{\ast 0}-\Sigma^{0}$ &  $\Xi^{\ast 0}-\Xi^{0}$  \\ \hline
$\delta M_{B^{\ast}} - \delta M_B$ [MeV] 
& 0 & $4.6 \times 10^2$ & $1.6 \times 10^2$ & $2.2 \times 10^2$ \\
$\Delta m_{B^\ast-B}|_\mathrm{theory}$ [MeV] 
& $3.7 \times 10^2$ & $8.2 \times 10^2$ & $5.2 \times 10^2$ &$5.9 \times 10^2$ \\
$\Delta m_{B^\ast-B}|_\mathrm{exp}$ [MeV] 
& $2.3 \times 10^2$ & $2.9 \times 10^2$ & $1.9 \times 10^2$ & $2.2 \times 10^2$ \\
\hline
\end{tabular}
\end{center}
\end{table}%

Finally, we focus on octet-decuplet mass difference.
A problem is known that the overall magnitude of the baryon spectra 
is not satisfactorily reproduced in the Sakai-Sugimoto model,
once we use  \eqref{values_of_M_KK_and_kappa}
\cite{HataMurata,HSSY}.
For instance, in the case of massless three flavors,
the mass difference between octet and decuplet baryons 
was calculated in \cite{HataMurata} as
\begin{eqnarray}
M_\mathbf{10} - M_\mathbf{8} = 0.386208 \times M_\mathrm{KK} 
\sim 3.7 \times 10^2 \mathrm{[MeV]} \equiv M_0,
\end{eqnarray}
where $M_\mathrm{KK}=949$ MeV is used.
However, the observed mass difference between the average value of octet baryons and 
that of decuplet ones is around 230 MeV.

Once we include the quark mass, this mass difference is modified 
as shown in Table~\ref{mass_splitting_8_10},
where  $B$ and $B^\ast$ label octet and decuplet baryons, respectively.
$\delta M_{B^{\ast}} - \delta M_B$ is the contribution from the nonzero quark masses, 
and the octet-decuplet difference is evaluated as
\begin{eqnarray}
\Delta m_{B^\ast-B}|_\mathrm{theory} = M_0 + \delta M_{B^{\ast}} - \delta M_B.
\end{eqnarray}
From Table~\ref{mass_splitting_8_10}, we observe that, at the leading order in quark masses, 
the inclusion of quark masses does not improve the situation of this baryon mass problem.
We believe higher order corrections other than the quark masses are relevant for resolving this problem.

\section*{Acknowledgments}

K.H.~and N.I.~would like to thank Kavli Institute for Theoretical Physics at
Santa Barbara. 
K.H.~also thanks
Yukawa Institute for Theoretical Physics at Kyoto University, at
which this topic was discussed during the workshop YITP-W-09-04 on 
``Development of Quantum Field Theory and String Theory.''
N.I.~also thanks RIKEN.
K.H.~is partly supported by
the Japan Ministry of Education, Culture, Sports, Science and
Technology.
T.I.~is partly supported by JSPS Research Fellowships for Young Scientists.

\providecommand{\href}[2]{#2}\begingroup\raggedright\endgroup

\end{document}